\documentclass[12pt]{article}
\newcommand{\be}{\begin{equation}}
\newcommand{\beq}{\begin{equation}}
\newcommand{\ee}{\end{equation}}
\newcommand{\eq}{\end{equation}}
\newcommand{\bea}{\begin{eqnarray}}
\newcommand{\eea}{\end{eqnarray}}
\newcommand{\vcut}{\delta x_\lambda}
\newcommand{\cut}{x_\lambda}
\newcommand{\step}{\Delta x}
\input feynman

\begin{document}

\vspace*{-1.5cm}

\begin {flushright}

LPT-ORSAY 00-76

\end{flushright}
\begin{center}
\baselineskip=13pt

\vspace{.7CM}

{\Large \bf  Soft gluon cascades and BFKL equation \\}
\vskip1.7cm
{\Large Andrei Shuvaev}\\
\vskip.7cm
{\it St. Petersburg Nuclear Physics Institute \\
188350, Gatchina, St. Petersburg district, Russia}
\vskip.3cm
{\tt e-mail: shuvaev@thd.pnpi.spb.ru}

\vskip1.7cm
{\Large Samuel Wallon}\\
\vskip.7cm
{\it Laboratoire de Physique Th\'eorique\,\footnote{Unit{\'e} mixte 8627 du CNRS} \\
Universit\'e Paris XI, Centre d'Orsay, b\^at. 211 \\
91405 Orsay Cedex, France}
\vskip.3cm
{\tt e-mail: Samuel.Wallon@th.u-psud.fr}

\end{center}
\vspace*{1.5cm}

\begin{abstract}
In this paper we deal with high energy scattering in the Regge limit,
 using a soft cascade approach.
We derive an evolution equation for the gluon density in soft gluons cascades
in the leading logarithmic
appro\-ximation of perturbative QCD, 
and show that this equation reproduces BFKL equation in the forward case.
The whole cascade is equivalent to a single gluon whose self-energy
is responsible for gluon reggeization. 
The same type of equation is obtained for the QED case.
\end{abstract}

\mbox{}
\thispagestyle{empty}
\newpage
\mbox{}
\thispagestyle{empty}
\newpage
\setcounter{equation}{0}
\setcounter{page}{1}
\setcounter{footnote}{0}
\setcounter{figure}{0}

\section{Introduction}
\label{introduction}

Investigation of hard scattering amplitude in the kinematics where
invariant energy $s$ is large while transfered momentum $t$ is fixed is
an important problem of QCD. The solution of this problem in the
Leading Logarithmic Approximation (LLA), collecting terms of the form
$(\alpha_S \, \ln s)^n\, ,$ was proposed about twenty five years ago,
in the BFKL equation \cite{BFKL}. These original papers were based on
an effective triple gluons vertex and a bootstrap condition in
$t-$channel.
This approach shew the relation between perturbative QCD and reggeon
calculus which was proposed a decade before. The main feature of this work
is the reggeization of gluon, which appears not to be elementary
but composite, as being a pole in complex momentum plane, with color octet
quantum number. Another
important result is that Pomeron occurs as a bound state of two
reggeized gluons in singlet color channel.
Although BFKL equation looks like a ladder type equation, it effectively
sums up an infinite number of $t-$channel gluons.

A few years ago a dipole picture was proposed
\cite{mueller94,muellerpatel,muellerunitarite,muellerchen,nik,nikital}. It is
based on subsequent emissions of soft gluons, which iterates
in the large $N_c$ limit. In this approach elementary degrees of freedom
are made of colored dipole. The equation for dipole density in this soft cascade
turns out to be identical to the BFKL equation, although it was
found in the multicolor limit, while original BFKL was derived for
finite $N_c.$

The aim of this paper is to show that BFKL dynamics can be simply
re\-co\-ve\-red in the forward limit
only using this classical soft emission vertex used in the dipole
model \cite{Navelet:1997tx}. We deal with gluons at finite $N_c$ rather than with dipoles
at large $N_c.$ We obtain BFKL equation
as an evolution equation for the gluon density in the cascade. This
equation is similar to DGLAP \cite{GL,AP,GAltarelli,dok} except
that the evolution variable is $\ln 1/x$
rather than $\ln Q^2\,.$ Some related approaches were elaborated in
 \cite{Korchemskaya:1996je} and \cite{Ball:1997vf}.

Our paper goes as follows.
 After preliminary recalling linear classical
cascades properties, we generalize them in order to incorporate
non abelian effects in the LLA approximation.
The main idea is that despite the complicated internal structure of the
cascade,
the amplitude of soft emission is the same as the one off a single gluon.
Using this idea we
rederive BFKL equation for forward
case.  The same
technique, when applied to QED, gives the same  type of equation.

\section{Classical cascade}
\label{classical}

Consider the scattering of two particles.
The asymptotic behavior of the
cross section is determined by the parton cascade from incoming particles
or, in other words, by their wavefunction on the light cone.
For large invariant energy $s$ (the so-called
Regge limit), the amplitude is dominated by
$t$-channel gluons rather than quarks, because of their largest
spin. As a consequence gluons dominate in every channel, and quark
contributions will be neglected below.

We shall consider a semi-hard region where gluons can be treated to be
soft with respect to the large invariant energy $s,$ although still
perturbative. It is known that in any massless field theory scattering
amplitudes are dominated by IR contributions of two types, namely
collinear and
soft singularities. This semi-hard region reveals soft gluons
contributions, which sum up into LLA.

The physics
behind soft resummation is that the field of an
ultrarelativistic source can be treated as a state of almost real particles
while the vertex for real or virtual soft gluon emission can be taken
to be the same.

Let us denote
$p_A$ and $p_B$ the momenta of colliding particles $A$ and $B.$
Neglecting their masses in the high energy limit we can take
$p_A^2\,=\,p_B^2\,=0$ so that the invariant energy reads $s\,=\,2p_A\cdot p_B$.
Vectors $p_A$, $p_B$ can be used for the Sudakov decomposition of any
vector as
$k=\alpha p_A + \beta p_B + k_\perp$.

Each particle develops a parton cascade, so the scattering of the incoming
particles $A$ and $B$ could be in principle reduced to elementary parton
processes. However the complete description of the cascade is a very
complicated pro\-blem involving all perturbative as well as nonperturbative
effects. There are certain limits, where the cascade looks more simple.
A well known example is  deep inelastic scattering. For large
virtuality $Q^2$ the parton density obeys the DGLAP evolution equation
collecting the powers of $\alpha_S \, \ln Q^2$. Although predictions
based on the DGLAP
evolution yield good agreement with the present experimental data,
this approach fails for parametrically small Bjorken variables $x$, when
the powers of $\alpha_S \, \ln 1/x$ dominate. This is the region of Regge kinematics
where the total invariant energy is much greater than the other invariants,
including the virtuality of the deeply virtual photon. The leading $\ln 1/x$
behavior of the hard amplitude is given in this region by the BFKL theory.

An important feature of the Regge kinematics is that the partons are soft there
in the sense that the main contribution appears for small longitudinal
variables, $\alpha, \beta \ll 1$.  Only soft
emissions from the external incoming or outgoing particles should be
taken into account in this approximation. Since the momenta of
 all incident external
particles are supposed to be along $p_A$ or $p_B$ directions,
two peculiar gauges are of special interest, namely the gauges
$p_A \cdot A=0 $ and $p_B \cdot A =0.$ These gauges suppress soft emission
 from $p_A$ and $p_B$ lines res\-pectively. In the following we shall deal
with emission from $p_B$ line in the gauge $p_A \cdot A=0.$

The vertex for emitting a soft particle
is rather universal and determined by the classical current proportional
to the momentum of the source. It can be easily obtained, for example, from
the triple gluon vertex in the soft  limit.  Consider emission off the
incoming particle $B,$ as illustrated in Fig.\,\ref{Figsoftvertex}.

\vspace{0.5cm}
\begin{figure}[htp]
\begin{picture}(5000,5000)(-12000,2000)
\global\Xone=1500
\startphantom
\drawline\fermion[\E\REG](500,1000)[\Xone]
\global\double\Xone
\stopphantom
\put(\pfrontx,\pfronty){\circle{\Xone}}
\drawline\gluon[\E\REG](\pbackx,\pbacky)[8]
\drawarrow[\W\ATBASE](8500,\pbacky)
\global\advance \pfrontx by 1500
\global\advance \pfronty by -2000
\global\advance \pbackx by -1500
\global\advance \pbacky by -2000
\put(\pbackx,\pbacky){$\! \! \! \! \! p_B, \,a$}
\put(\pfrontx,\pfronty){$a^\prime$}
\drawline\gluon[\NE\REG](\pmidx,\pmidy)[4]
\drawarrow[\E\ATBASE](\pbackx,\pbacky)
\global \advance \pbacky by 500
\global\advance \pmidx by 1500
\global\advance \pmidy by -600
\put(\pbackx,\pbacky){$c,\,k$}
\end{picture}

\vspace{0.8cm}

\caption{Soft vertex for incoming line. The blob symbolizes any
amplitude connected to the incoming line.}
\label{Figsoftvertex}
\end{figure}
This incoming line is attached to an amplitude
(shown as a blob in Fig.\,\ref{Figsoftvertex}) whose peculiar form plays
no role so long as we are dealing with soft emission. The soft vertex reads
\begin{equation}
\label{sv}
\Gamma^\lambda_c(k)\,=\,g\,
\frac{p_B\cdot \varepsilon^\lambda(k)}{p_B k -i\delta}\,T_c\,=\,
g\,\frac{p_B\cdot \varepsilon^\lambda(k)}{\alpha \frac s2 -i\delta}\,T_c,
\end{equation}
where $\varepsilon^\lambda(k)$ is the helicity vector of the soft emitted or
absorbed gluon carrying momentum $k$ and color index $c$.
The matrix $T^c$ depends on the representation of the color group.
For gluon like object it is expressed through the structure constants,
$T^c_{a^\prime a} = i f_{a^\prime c a}$, the indices $c,a^\prime,a$ being
in the adjoint representation.

\vspace{0.5cm}

\begin{figure}[htb]
\begin{picture}(5000,5000)(-6000,2000)
\thinlines
\startphantom
\drawline\fermion[\E\REG](500,1000)[1430]
\stopphantom
\put(\pfrontx,\pfronty){\circle{3000}}
\drawline\fermion[\E\REG](\pbackx,\pbacky)[2560]
\global\Xone=\pfrontx
\global\advance\pmidy by -50
\thinlines
\global\advance \pfrontx by 500
\global\advance \pfronty by -1500
\put(\pfrontx,\pfronty){$a^\prime$}
\drawline\gluon[\NE\REG](\pbackx,\pbacky)[4]
\drawarrow[\E\ATBASE](\pbackx,\pbacky)
\global \advance \pbacky by 500
\global\advance \pmidx by 1000
\global\advance \pmidy by -600
\put(\pbackx,\pbacky){$c_n, \, k_n$}
\global\advance \pmidx by 3000
\put(\pmidx,\pmidy){$\cdot\; \cdot\; \cdot$}
\thinlines
\drawline\fermion[\E\REG](\gluonfrontx,\gluonfronty)[10000]
\thinlines
\drawline\gluon[\NE\REG](\pbackx,\pbacky)[4]
\drawarrow[\E\ATBASE](\pbackx,\pbacky)
\global \advance \pbacky by 500
\global\advance \pmidx by 1000
\global\advance \pmidy by -600
\put(\pbackx,\pbacky){$c_2, \, k_2$}
\THINLINES
\drawline\fermion[\E\REG](\fermionbackx,\fermionbacky)[6000]
\thinlines
\drawline\gluon[\NE\REG](\pbackx,\pbacky)[4]
\drawarrow[\E\ATBASE](\pbackx,\pbacky)
\global \advance \pbacky by 500
\global\advance \pmidx by 1000
\global\advance \pmidy by -600
\put(\pbackx,\pbacky){$c_1,\,k_1$}
\thinlines
\drawline\fermion[\E\REG](\fermionbackx,\fermionbacky)[6000]
\thinlines
\global\advance \pbackx by -1000
\global\advance \pbacky by -1300
\put(22500,\pbacky){$p_B, \,a$}
\thinlines
\global\Yone=\Xone
\global\advance\Yone by -26000
\global\advance\Yone by -500
\global\negate\Yone

\drawline\fermion[\E\REG](\Xone,900)[\Yone]
\drawarrow[\W\ATBASE](22500,950)
\end{picture}
\vskip 1 cm
\caption{Amplitude for $n$ gluon emission or absorption off an incoming gluon-like
object, represented as two thin lines.}
\label{Figwilson}
\end{figure}

\vskip 0.3cm

The amplitude for emission or absorption of $n$ gluons (see
Fig.\,\ref{Figwilson}) is given by the product of the elementary vertices,
\begin{equation}
\label{nga}
\Gamma^{\mu_n,\ldots \mu_1}_{c_n,\ldots c_1}(k_1,\ldots
k_n)\,A_{\mu_n,\,c_n}(k_n) \cdots \,A_{\mu_1,\,c_1}(k_1)\,=\,(T^{c_n}\cdots
T^{c_1})_{a' a}
\times
\end{equation}
$$
\times
g\frac{p_B\cdot A_{c_n}(k_n)}{(\alpha_1+\alpha_2+\ldots
+\alpha_n)\frac s2 - i\delta}\, \cdots
g\frac{p_B\cdot A_{c_2}(k_2)}{(\alpha_1+\alpha_2)\frac s2 -
i\delta}\,
\,g\frac{p_B\cdot A_{c_1}(k_1)}{\alpha_1\frac s2 - i\delta}\,
.
$$
Here the field $A$ is the asymptotic free field
$$
A_{\mu,\,c}(x)= \frac{1}{(2 \pi)^{3/2}} \int \frac{d^3 k}{2 k_0}
\left [ e^{ik x} \varepsilon_{\mu}^{\lambda}(k) \, a^+_{\lambda,\,c}(k)
+ e^{-ik x} \varepsilon_{\mu}^{\lambda}(k) \, a_{\lambda,\,c}(k) \right ] \,,
$$
whose positive and negative frequency part corresponds respectively to
emission and absorption of gluons.
Using the relation
$$
\frac 1 {\alpha_1\frac s2 - i\delta}\,=\,i\int_{-\infty}^\infty dz\,
e^{i \alpha \frac s2 z} \theta(-z)
$$
we get
$$
\int d^4k_1\cdots d^4k_n\,\Gamma^{\mu_n,\ldots \mu_1}_{c_n,\ldots c_1}(k_1,
\ldots k_n)\,A_{\mu_n,\,c_n}(k_n)\, \cdots \,A_{\mu_1,\,c_1}(k_1)
$$
$$
=\,\frac{(ig)^n}{n!} \int dz_1\cdots dz_n\,{\rm P}\bigl[n_B\cdot A(z_n n_B)\,\cdots\,
n_B\cdot A(z_1 n_B)\bigr]_{a^\prime a}\,,
$$
with
$$
A_\mu(x)\,=\,\int d^4k\,e^{ikx}A_{\mu,\,c}(k)\,T^c
$$
and where the symbol P means the conventional ordering of the fields. The sum over the
arbitrary number of soft gluons is given by a P-exponent along the momentum
direction of the parent particle:
\bea
\label{wilson}
&&\sum_n
\int d^4k_1\cdots d^4k_n\,\Gamma^{\mu_n,\ldots \mu_1}_{c_n,\ldots c_1}(k_1,
\ldots k_n)\,A_{\mu_n,\,c_n}(k_n)\, \cdots \,A_{\mu_1,\,c_1}(k_1)\,
\nonumber \\
&& \hspace{2cm} =\,{\rm P}\exp\bigg[ig \int_{-\infty}^0 dz\,n_B\cdot A(z n_B)\bigg].
\eea
A similar computation for outgoing line results in replacing in the
previous formula $\int_{-\infty}^0$ with $\int_0^\infty.$
These formulae are nothing more than the Wilson line taken on the
trajectory along the momentum of the incoming (resp. outgoing) particle.

\section{Inclusion of interaction among emitted particles}
\label{interaction}

The previous computation only takes into account emissions from the
pa\-rent line which is the incoming line (resp. outgoing).
The above formulae are strictly speaking only valid for QED. They do
not take into account the interaction between emitted particles. In QCD,
in the soft approximation, these corrections correspond to subsequent
decays of the emitted gluons. Indeed in this approximation only end
lines emissions contribute, and thus
to get the total cascade tree-level amplitude each field
in the P-exponent has to be replaced by the P-exponent itself.
Generally this turns into a complicated non-linear equation.
Another problem is to incorporate loop corrections. This second problem will
be considered in section \ref{virtual}.

The first problem, the partons subsequent decays, simplifies in the Regge
kinematics implying the longitudinal momenta to be small in
the formula (\ref{nga}), $\alpha,\beta \ll 1$, while
\begin{equation}
\label{kin}
\alpha s,\beta s \gg k_\perp^2, \qquad \alpha\beta s\sim k_\perp^2\,,
\end{equation}
where
$k_\perp$ is a typical transverse momentum scale. This kinematics
ensures the soft vertex to be of the form (\ref{sv}) regardless from where
a particular gluon is emitted off or where it is absorbed in.
Indeed, the emission of a gluon with momentum $k_2=\alpha p_A+\beta
p_B+k_{2\perp}$ off the  parent gluon carrying momentum $k_1=a p_A+b p_B
+k_{1\perp}$ is again given by a soft vertex similar to (\ref{sv})
\begin{equation}
\label{softvertex}
\Gamma^\lambda_c(k_2)\,=\,g\,\frac{k_1\cdot
\varepsilon^\lambda(k_2)}{k_1\cdot k_2}\,T_c\,.
\end{equation}
In the light cone gauge $p_A\cdot A(x)=0$, the polarization vector reads
\begin{equation}
\label{eps}
\varepsilon^{\lambda}_{\mu}(k)\,=\,\varepsilon^{\lambda}_A(k)
\cdot p_{A\mu}\,+\,\varepsilon^{\lambda}_{\perp \mu}(k)
\end{equation}
with
$$
\sum_\mu\varepsilon^{\lambda}_{\perp \mu}(k)\cdot
\varepsilon^{\lambda^\prime}_{\perp \mu}(k)\,=\,
\delta^{\lambda \lambda^\prime}.
$$
The emitted gluons are on-shell. Anyway, the gluons
which dominate the amplitude in the Regge limit are
soft and thus quasi-real, therefore the on-shell condition is unessential.
The transversality condition for these quasi-real gluons reads
$k\cdot\varepsilon = 0$, which
implies for the helicity vector (\ref{eps})
\begin{equation}
\label{keperp}
\varepsilon^{\lambda}_A(k)\,=\,2\,
\frac{k_\perp \cdot \varepsilon^{\lambda}_{\perp }(k)}{\beta s}\,.
\end{equation}
In this soft gluon approximation, $\beta \ll b$ and assuming
the typical transverse momenta to
be of the same order, $k_{1\perp}^2 \sim k_{2\perp}^2$,
the numerator of the soft vertex is approximated as
$$
k_1\cdot \varepsilon^\lambda(k_2)\,=\,b \,\varepsilon^\lambda_A\,
\frac s2-k_{1\perp} \varepsilon^\lambda_\perp\,\approx \, b \,p_B \cdot
\varepsilon^\lambda(k_2).
$$
Using the mass shell conditions, $a=k_{1\perp}^2/b s$,
$\alpha=k_{2\perp}^2/\beta s,$ the denominator also simplifies as
$$
k_1\cdot k_2\,=\,(a \,\beta + b \,\alpha)\,\frac s2 \,-\,k_{1\perp}\cdot
k_{2\perp}\,\approx \, b \, \alpha\,\frac s2 \,.
$$
It follows that the vertex (\ref{softvertex}) can be written as
\begin{equation}
\label{vertexk2}
\Gamma^\lambda_c(k_2)\,=\,g
\frac{p_B\cdot \varepsilon^{\lambda}(k_2)}{p_B\cdot k_2} T_c \,,
\end{equation}
which has exactly the universal form (\ref{sv}). It confirms the
physical picture that soft emission is determined by the total current
of the source rather than by its internal structure.
The vector $p_B$ is the momentum of the source.

With this universal vertex combined with Jacobi identity for the color
matrices $T^c$ the emission of the new soft particle from the "ends"
$a$ and $c$ in the Fig.\,1 looks like if it were effectively emitted off
the line $a^\prime$ at the left from the particle $c$.
This is illustrated in Fig.\,\ref{jacobi}.

\vskip 1cm
\begin{figure}
\begin{picture}(5000,19000)(-4000,-11500)
\global\Xsix=-1400
\global\Xseven=2800
\thinlines
\drawline\fermion[\E\REG](500,1000)[3000]
\global\Xfive=\pfrontx
\global\advance\Xfive by \Xsix
\put(\Xfive,\pfronty){\circle{3000}}
\thinlines
\global\advance \pfrontx by 1000
\global\advance \pfronty by \Xsix
\put(\pfrontx,\pfronty){$a^\prime$}
\drawline\gluon[\NE\REG](\pbackx,\pbacky)[4]
\drawarrow[\E\ATBASE](\pbackx,\pbacky)
\global \advance \pbacky by 500
\global\advance \pmidx by 1000
\global\advance \pmidy by -600
\put(\pbackx,\pbacky){$c_1, k_1$}
\thinlines
\drawline\fermion[\E\REG](\fermionbackx,\fermionbacky)[6000]
\thinlines
\drawline\gluon[\NE\REG](\pbackx,\pbacky)[4]
\drawarrow[\E\ATBASE](\pbackx,\pbacky)
\global \advance \pbacky by 500
\global\advance \pmidx by 1000
\global\advance \pmidy by -600
\put(\pbackx,\pbacky){$c_2, k_2$}
\THINLINES
\drawline\fermion[\E\REG](\fermionbackx,\fermionbacky)[5000]
\global\Xone=\fermionbackx
\global\advance\Xone by -500
\thinlines
\global\advance \pbackx by -1000
\global\advance \pbacky by -1300
\put(\pbackx,\pbacky){$a$}
\startphantom
\drawline\fermion[\E\REG](\fermionbackx,\fermionbacky)[2000]
\stopphantom
\put(\pbackx,700){$+$}
\global\Xeight=2500
\global\advance \Xeight by \Xseven
\startphantom
\drawline\fermion[\E\REG](\fermionbackx,\fermionbacky)[\Xeight]
\stopphantom
\thinlines
\drawline\fermion[\E\REG](\pbackx,\pbacky)[8000]
\global\Xfive=\pfrontx
\global\advance\Xfive by \Xsix
\put(\Xfive,\pbacky){\circle{3000}}
\global\Xtwo=\pfrontx
\global\advance \pfrontx by 1000
\global\advance \pfronty by -1500
\put(\pfrontx,\pfronty){$a^\prime$}
\global\advance \pbackx by -1000
\global\advance \pbacky by -1500
\put(\pbackx,\pbacky){$a$}
\drawvertex\gluon[\NE 3](\pmidx,\pmidy)[2]
\drawarrow[\N\ATBASE](\vertextwox,\vertextwoy)
\drawarrow[\E\ATBASE](\vertexthreex,\vertexthreey)
\global\advance\vertexthreex by 800
\put(\vertexthreex,\vertexthreey){$c_1, k_1$}
\global\advance\vertextwox by -3000
\put(\vertextwox,\vertextwoy){$c_2, k_2$}
\thinlines
\drawline\fermion[\E\REG](500,900)[\Xone]
\drawline\fermion[\E\REG](\Xtwo,900)[8000]
\startphantom
\drawline\fermion[\S\REG](500,1000)[10000]
\stopphantom
\put(\pbackx,-9300){$=$}
\startphantom
\drawline\fermion[\E\REG](\pbackx,\pbacky)[\Xeight]
\stopphantom
\thinlines
\drawline\fermion[\E\REG](\pbackx,\pbacky)[4000]
\global\Xfive=\pfrontx
\global\advance\Xfive by \Xsix
\put(\Xfive,\pfronty){\circle{3000}}
\global\Xone=\pfrontx
\global\Xtwo=\pfronty
\global\advance\Xtwo by -100
\thinlines
\global\advance \pfrontx by 1000
\global\advance \pfronty by \Xsix
\put(\pfrontx,\pfronty){$a^\prime$}
\drawline\gluon[\NE\REG](\pbackx,\pbacky)[4]
\drawarrow[\E\ATBASE](\pbackx,\pbacky)
\global \advance \pbacky by 500
\global\advance \pmidx by 1000
\global\advance \pmidy by -600
\put(\pbackx,\pbacky){$c_2, k_2$}
\thinlines
\drawline\fermion[\E\REG](\fermionbackx,\fermionbacky)[6000]
\thinlines
\drawline\gluon[\NE\REG](\pbackx,\pbacky)[4]
\drawarrow[\E\ATBASE](\pbackx,\pbacky)
\global \advance \pbacky by 500
\global\advance \pmidx by 1000
\global\advance \pmidy by -600
\put(\pbackx,\pbacky){$c_1, k_1$}
\THINLINES
\drawline\fermion[\E\REG](\fermionbackx,\fermionbacky)[5000]
\global\Xthree=\fermionbackx
\global\negate\Xone
\global\advance\Xthree by \Xone
\global\negate\Xone
\global\advance\Xone by -500
\thinlines
\global\advance \pbackx by -1000
\global\advance \pbacky by -1300
\put(\pbackx,\pbacky){$a$}
\global\negate \Xsix
\global\advance \Xfive by \Xsix
\drawline\fermion[\E\REG](\Xfive,\Xtwo)[\Xthree]
\end{picture}
\caption{Illustration of the soft gluon universality using Jacobi identity.}
\label{jacobi}
\end{figure}
\vskip 5cm
Since the emitted gluon is real, $k^2 = \alpha \, \beta \, s -
k_\perp^2=0,$
and thus, combining Eqs.(\ref{eps}) and (\ref{keperp}), the vertex
(\ref{vertexk2})
can be rewritten  in the form
\beq
\label{universal}
\Gamma^\lambda_c(k) = g \, T^c \, \frac{2 k_{\perp} \cdot \epsilon_\perp}{k_\perp^2}\,.
\eq
Repeating the new emissions from all "ends" in the same manner we come back
to the Fig.\,\ref{Figwilson}, where now the gluons are ordered according
to their $\beta$ value, the smaller $\beta$ being  on the left of
the largest ones.

Note that the decreasing order of $\beta$ variables,
\beq
\label{betaordering}
\beta_1\, \gg\, \beta_2\,\gg\,\cdots\,\,\gg\,\beta_n,
\eq
implies in the Regge kinematics increasing order of $\alpha$'s,
$$
\alpha_1\, \ll\, \alpha_2\,\ll\,\cdots\,\,\ll\,\alpha_n,
$$
so that $\alpha_1+\alpha_2\simeq \alpha_2$,
$\alpha_1+\cdots+\alpha_n\simeq \alpha_n$.
This property allows to present the resulting soft cascade tree amplitude
through an ordered exponent in momentum space,
$$
\Phi_{tree}\,=\,P_\beta\exp\left\{\int_{x_\lambda}^1d\beta\,
d^2k_\perp\,V(\beta,k_\perp)\right\},
$$
with
\beq
\label{defV1}
V(\beta,k_\perp)\,=\,V^+(\beta,k_\perp)\,+\,V^-(\beta,k_\perp)
\ee
and
\beq
\label{defV2}
V^\pm(\beta,k_\perp)\,=\,\frac g{(2\pi)^{\frac 32}}\,\frac{1}{2 \beta}
\frac{p_B \cdot \varepsilon^\lambda(\beta,k_\perp)}{p_B \cdot k}\,a_{\lambda,\,c}^\pm(\beta,k_\perp)
\,T^c.
\ee
In the gauge $p_A \cdot A =0,$ this equation reduces to
\beq
\label{V}
V^\pm(\beta,k_\perp)\,=\,\frac {2 \,g \, T^c}{(2\pi)^{\frac 32}}\,\frac{1}{2 \beta}
\frac{k_\perp \cdot \varepsilon^\lambda_\perp(\beta,k_\perp)}{k_\perp^2}\,a_{\lambda,\,c}^\pm(\beta,k_\perp)
\,
\ee
where we have used expression (\ref{universal}).
This expression incorporates both emission and absorption of the partons
described by the light cone creation, $a^+_{\lambda c}(\beta,q_\perp)$,
and annihilation, $a_{\lambda c}(\beta,q_\perp)
\equiv a^-_{\lambda c}(\beta,q_\perp)=
a^+_{\lambda c}(-\beta,-q_\perp)$, operators labelled by the
longitudinal momentum fraction $\beta \ge 0$, transverse momentum
$q_\perp$, polarization and color indices $\lambda$ and $c$, which satisfies
$$
\bigl[a_{\lambda c}(\beta,q_\perp)\,,
\,a^+_{\lambda^\prime c^\prime}(\beta^\prime,q_\perp^\prime)\bigr]\,=\,
2\beta\, \delta_{\lambda \lambda^\prime}\delta_{c c^\prime}\,
\delta(\beta-\beta^\prime)\delta^{(2)}(q_\perp-q_\perp^\prime).
$$

The ordering symbol $P_\beta$ means that the fields $A(\beta)$ with
the smallest $\beta$ value are on the left from the fields with
the largest ones. The minimal value $\beta=x_\lambda$ plays the role of
an infrared cut-off in the amplitude $\Phi_{tree}$.
Note that in the case of incoming line, the variable ``time'' $z$ in the
P-exponent (\ref{wilson}) is Fourier conjugated to the longitudinal Sudakov
variable $\alpha,$ and thus ordered oppositely. Since on the other hand,
for mass-shell particles $\alpha \sim 1/\beta,$ it turns out that $|z|
\sim \beta.$ Thus, P-exponent defined with respect to $z$ is in
accordance
with $P_\beta.$

When the cut-off value is lowered, that is when we take
$x_\lambda^\prime < x_\lambda$, it allows for the emission or absorption
of an extra soft particle whose longitudinal momentum lies in the interval
$x_\lambda^\prime < \beta < x_\lambda$. This yields the new amplitude
\begin{equation}
\label{tree}
\Phi_{tree}^\prime\,=\,\left[1 +
\int^{x_{\lambda}}_{x_{\lambda}-\delta x_{\lambda}}\,d \beta\,
\int d^2 k_\perp \,V(\beta,\,k_\perp)\right]\, \Phi_{tree}\,,
\end{equation}
the amplitude $\Phi_{tree}$ standing as the source for the new
soft particles. The whole tree amplitude can be symbolically presented
as an infinite product of elementary emissions or absorptions in
the infinitesimal intervals $\Delta x$,
\begin{equation}
\label{Ptr}
\Phi_{tree}\,=\,\prod\limits_{x_\lambda}^1
\left[1\,+\,\int^{x}_{x-\Delta x}\, d \beta \,\int d^2 k_\perp\,
V(\beta,\,k_\perp)\right].
\end{equation}

\section{Virtual corrections and evolution of the cascade wave function}
\label{virtual}

The previous form is convenient to include the virtual contributions. Besides
the amplitude to emit or absorb a real gluon at each elementary
step there is an amplitude for the case when the new phase volume in the longitudinal
space remains non-occupied. In other words it means that the emitted
gluon is reabsorbed at the same step, which corresponds to a loop correction. It
does not changes the number of particles, but it is responsible for
the renormalization of the whole state.

Despite the composite internal structure of the whole system of cascade
and source, within the soft
emission approach this system 
 looks like the source carrying given momentum
and color.  
The virtual correction results into self-energy insertion into the
source propagator. In particular, if the source is gluon like, then the full
system looks like a gluon.
In this case, the virtual correction results into self-energy insertion
in the gluon propagator. In LLA and for the cascade plus source
system of momentum
$q \simeq \, p_B+q_\perp$, the propagator of this system is modified as
$$
D_{\mu \nu}(q)\,=\,\frac 1{q^2}\,\left[\delta^{\mu \nu}\,-\,
\frac{q^\mu p_A^\nu + q^\nu p_A^\mu}
{p_A\cdot q}\right]\,\frac 1{1-\pi(x_\lambda,q_\perp^2)},
$$
with the function $\pi(x_\lambda,q_\perp^2)$ determining $z$-factor,
or normalization of the state. To keep the value of the norm fixed,
the cascade plus source
wavefunction is multiplied by $z_{[x_\lambda,1]}^{-1/2}(q_\perp)$, where
$q_\perp$ is the total transverse momentum of the system whose constituents
fill the interval of longitudinal momenta $[x_\lambda,1]$,
$$
z_{[x_\lambda,1]}(q_\perp)\,=\,\frac 1{1-\pi(x_\lambda,q_\perp^2)}.
$$
The contribution of the soft gluons appearing at the next evolution step
in the interval $[x_\lambda-\delta x_\lambda,x_\lambda]$ should
be compensated by the factor
$z_{[x_\lambda-\delta x_\lambda,x_\lambda]}^{-1/2}(q_\perp)$. Basically for
a finite interval one can write a loop expansion of the form
$$
z_{[x_\lambda-\vcut,\,x_\lambda]}(q_\perp)\,=\,1\,+\,
\sum_{n>1}z_n(q_\perp)\,\ln^n\frac{x_\lambda}{x_\lambda-\delta x_\lambda},
$$
only the first variation determining the evolution equation,
\begin{equation}
\label{zomega}
z_{[x_\lambda-\delta x_\lambda,x_\lambda]}^{-1/2}(q_\perp)\,=\,
1\,-\,\omega(q_\perp)\,\frac{\delta x_\lambda}{x_\lambda}\,+\,
O\left(\frac{\delta x_\lambda}{x_\lambda}\right)^2.
\end{equation}
In LLA function $\omega(q_\perp)$ is given by a one loop diagram
calculated in the axial gauge.

Before discussing an explicit form of the virtual contribution note that
the momentum $q_\perp$ in the argument of $\omega$ is the total
transverse momentum of the system of
 cascade and source, which  emit a soft particle as a whole.
Introducing the total momentum operator $\hat P_\perp,$ $\omega(\hat P_\perp)$
obviously gives $\omega(q_\perp)$ when acting on a whole system of
cascade 
plus source state of
total momentum $q_\perp.$
Thus, in a similar way as in the tree case (see Eq.(\ref{tree})), one
step of evolution reads
$$
\Phi(\cut-\vcut)\,=\,
\left[1 - \omega(\hat P_\perp)\,\frac{\vcut}{x} \,+\,
\int^{x}_{x-\vcut}\,d\beta \,\int d^2 k_\perp \,
V(\beta,\,k_\perp)\right]\,\Phi(\cut)\,. 
$$

The full cascade $S$ matrix can be symbolically written through the elementary steps
product\footnote{Here $\delta x_\lambda$ denotes the variation of the cut-off value
while $\Delta x$ is taken as a notation for an infinitisemal step in
the infinite product representation of the cascade $S$ matrix. Principally
one can put $\delta x_\lambda=\Delta x$.}
\begin{equation}
\label{prodPhi}
\Phi(\cut)\,=\,\prod\limits_{\cut}^1
\left[1 - \omega(\hat P_\perp)\,\frac{\step}{x} \,+\,
\int^{x}_{x-\step}\,d\beta \,\int d^2 k_\perp \,
V(\beta,\,k_\perp)\right],
\end{equation}
where the brakets are ordered from the smallest $\beta$'s values at the left
to the greatest at the right.
The operator
$$
\Phi^+(x_\lambda)\,=\,\prod\limits_{x_\lambda}^1
\left[1 - \omega(\hat P_\perp)\,\frac{\step}{x} \,+\,
\int^{x}_{x-\Delta x}\,d\beta\,\int d^2 k_\perp \,
V^+(\beta,\,k_\perp)\right]
$$
acting on the vacuum creates the soft constituents of the cascade,
while the operator
$$
\Phi^+(x_\lambda,p_\perp)\,=\,\frac 1{(2\pi)^2}\,
\int d^{\,2}z\,e^{-ip_\perp z}\,e^{i\hat P z}\Phi^+(x_\lambda)\,
e^{-i\hat P z}
$$
picks out the components with a fixed transverse momentum.
Let us introduce the operator $b_{\sigma,a}^+(p_B,\,q_\perp)$ for 
creating the bare source,
having transverse momentum $q_\perp$, helicity and color indices
$\sigma$, $a$, 
and denote this state as $|p_B,q_\perp,a,\sigma\rangle_s\,=\, b_{\sigma,a}^+(p_B,q_\perp)|0\rangle.$
Then the state
$$
|\,p_B,x_\lambda, q_\perp,a,\sigma\,\rangle\,=\,\int d^2p_\perp\,
\Phi^+(x_\lambda,\,p_\perp)\,b_{\sigma,a}^+(p_B,\,q_\perp-p_\perp)|\,0\,\rangle
$$
describes the system of cascade plus source
 with given total transverse momentum $q_\perp$,
 cut-off value $x_\lambda,$ and source helicity and color. The total
 tranverse
momentum of the cascade is given by $p_\perp.$ The operator $\hat P_\perp$
acting on the right results at every step into the transverse momentum of
the total system at the previous cut-off value.

Changing $x_\lambda$ we evidently have for the variation  of the
cascade
plus source
state
\begin{eqnarray}
\label{varst}
\delta |\,p_B,x_\lambda, q_\perp,a,\sigma\,\rangle\,
&=&\,-\frac{\delta x_{\lambda}}{x_{\lambda}}\,\omega(q_\perp)\,
|\,p_B,\,q_\perp,\,x_\lambda,\,a,\sigma \,\rangle \\
&+&\,
\int\limits_{x_\lambda-\delta x_{\lambda}}^{x_\lambda}d\beta\,
\int d^2k_\perp\,V^+(\beta,k_\perp)\,
|\,p_B,\,(q-k)_\perp,\,x_\lambda,\,a,\sigma \,\rangle.
\nonumber
\end{eqnarray}

Let $Q$ be an operator, which probes the parton
distribution in the cascade,  for instance,\footnote{In principle the operator $Q$ could probe both the cascade content and the
source,
without modifying the following discussion. One could add to $Q$ a term
$$
1/2 \int d^2 l_\perp f^\prime (l_\perp) b_{\sigma, c}^+(p_B,l_\perp) b_{\sigma,
c}(p_B,l_\perp)
$$
acting on the source. If the source is gluon-like this is the same as to
include
in the weight function $f(x,l_\perp)$  in Eq.(\ref{defQ}) a
 term proportional to
$\delta(1-x)$.}
\beq
\label{defQ}
Q\,=\,\int_{x_Q}^1 \frac{dx}{2x}d^2l_\perp\,f(x,l_\perp)\,
a^+_{\sigma c}(x,l_\perp)\,a_{\sigma c}(x,l_\perp)
\eq
with a weight function $f(x,l_\perp)$. The average value is generally
expressed through the density function $n_f$, depending on the cascade
total transverse momentum and cut-off,
$$
\langle \,p_B,x_\lambda, q_\perp,a,\sigma\,|\,Q\,
|\,p_B,x_\lambda, q_\perp^\prime,a^\prime,\sigma^\prime\,\rangle\,=\,
\delta_{a, a^\prime}\delta_{\sigma,\sigma^\prime}
\delta^{(2)}(q_\perp - q_\perp^\prime)\,n_f(x_\lambda,q_\perp).
$$
Its evolution with $x_\lambda$ value is determined by the variation of
the state (\ref{varst}). If we suppose that the operator $Q$ has a natural
cut-off $x_Q \gg x_\lambda$, then the new emitted gluon does not affect
the operator vertex (the operators $a_{\lambda,c}(\beta,k_\perp)$ commute
with $Q$ for $x_\lambda-\delta x_{\lambda}<\beta<x_\lambda$) and
the variation of the matrix element decays into the sum
$$
\Delta n_f\,=\,\Delta_1 n_f\,+\,\Delta_2 n_f,
$$
as illustrated in Fig.\,\ref{Figvariationhomogeneous}.
\begin{figure}[htp]
\begin{picture}(15000,5000)
\global \Xone = 500                
\global \Yone = \Xone
\global \multiply \Yone by 2       
\global \Xtwo = 1300			 
\global \Xthree = \Xtwo
\negate \Xone
\global\advance \Xthree  by \Xone          
\negate \Xone
\global \Xfour = 300			   
\global \Xfive = 1100			   
\global \Ytwo = -250
\global \Ythree = -100		           
\global \Yfour = \Ythree
\global \multiply \Yfour by 2
\global \Xeight = 5000		
\global \Yeight = 500
\THICKLINES
\drawline\fermion[\E\REG](\Xeight,\Yeight)[\Xtwo]
\THINLINES
\global\advance\fermionbacky by \Ythree
\drawloop\gluon[\N 5](\pbackx,\pbacky)
\put(\loopmidx,\fermionbacky){\circle{\Yone}}
\negate \Xone
\global\advance\loopmidx by \Xone			
\negate \fermionbackx
\global\advance\loopmidx by \fermionbackx
\Xseven = \loopmidx
\advance \Xseven by \Xtwo				
\negate \Xone
\THICKLINES
\drawline\fermion[\E\REG](\loopfrontx,\loopfronty)[\loopmidx]
\drawline\fermion[\W\REG](\gluonbackx,\gluonbacky)[\fermionlengthx]
\drawline\fermion[\E\REG](\gluonbackx,\gluonbacky)[\Xtwo]
\advance\fermionbacky by \Yfour
\drawline\fermion[\E\REG](\Xeight,\fermionbacky)[\Xseven]
\advance\fermionbackx by \Yone
\drawline\fermion[\E\REG](\fermionbackx,\fermionbacky)[\Xseven]
\THINLINES
\startphantom
\drawline\fermion[\E\REG](\pbackx,\pbacky)[\Xfour]
\stopphantom
\global \Ysix = \Yeight
\global \advance\Ysix \Ytwo
\put(\pbackx,100){$+$}
\startphantom
\drawline\fermion[\E\REG](\pbackx,\pbacky)[\Xfive]
\stopphantom
\THICKLINES
\drawline\fermion[\E\REG](\pbackx,\Yeight)[\Xtwo]
\Xseven = \fermionfrontx			
\THINLINES
\drawloop\gluon[\N 5](\pbackx,\pbacky)
\negate\loopfrontx
\global\advance\loopbackx by \loopfrontx
\THICKLINES
\drawline\fermion[\E\REG](\fermionbackx,\pbacky)[\loopbackx]
\drawline\fermion[\E\REG](\fermionbackx,\fermionbacky)[\Xthree]
\negate \fermionbackx
\advance \Xseven by \fermionbackx
\negate \fermionbackx
\negate \Xseven			
\global\advance\fermionbackx by \Xone
\global\advance\fermionbacky by \Ythree
\THINLINES
\put(\fermionbackx,\fermionbacky){\circle{\Yone}}
\THICKLINES
\global\advance\fermionbackx by \Xone
\negate\Ythree
\advance\fermionbacky by \Ythree
\negate\Ythree
\drawline\fermion[\E\REG](\fermionbackx,\fermionbacky)[\Xthree]
\Yseven = \fermionbackx
\advance \fermionbacky by \Yfour
\drawline\fermion[\W\REG](\fermionbackx,\fermionbacky)[\Xthree]
\negate \Yone
\advance \fermionbackx by \Yone
\negate \Yone
\drawline\fermion[\W\REG](\fermionbackx,\fermionbacky)[\Xseven]
\THINLINES
\pbackx = \Yseven
\startphantom
\drawline\fermion[\E\REG](\pbackx,\pbacky)[\Xfour]
\stopphantom
\global \Ysix = \Yeight
\global \advance\Ysix \Ytwo					

\put(\pbackx,100){$+$}
\startphantom
\drawline\fermion[\E\REG](\pbackx,\pbacky)[\Xfive]
\stopphantom
\THICKLINES
\drawline\fermion[\E\REG](\pbackx,\Yeight)[\Xthree]
\Yseven = \fermionfrontx			
\global\advance\fermionbackx by \Xone
\global\advance\fermionbacky by \Ythree
\THINLINES
\put(\fermionbackx,\fermionbacky){\circle{\Yone}}
\THICKLINES
\global\advance\fermionbackx by \Xone
\Xseven = \fermionbackx					
\negate\Ythree
\advance\fermionbacky by \Ythree
\negate\Ythree
\drawline\fermion[\E\REG](\fermionbackx,\fermionbacky)[\Xthree]
\THINLINES
\drawloop\gluon[\N 5](\pbackx,\pbacky)
\negate\loopfrontx
\global\advance\loopbackx by \loopfrontx
\THICKLINES
\drawline\fermion[\E\REG](\fermionbackx,\fermionbacky)[\loopbackx]
\drawline\fermion[\E\REG](\fermionbackx,\fermionbacky)[\Xthree]
\negate \fermionbackx
\advance \Xseven by \fermionbackx
\negate \Xseven
\advance \fermionbacky by \Yfour
\drawline\fermion[\E\REG](\Yseven,\fermionbacky)[\Xthree]
\advance \fermionbackx by \Yone
\drawline\fermion[\E\REG](\fermionbackx,\fermionbacky)[\Xseven]
\THINLINES
\startphantom
\drawline\fermion[\E\REG](\pbackx,\pbacky)[\Xfour]
\stopphantom
\global \Ysix = \Yeight
\global \advance\Ysix \Ytwo					

\end{picture}
\vskip 0.5cm
\caption{Variation of the density with respect to the cut-off, in the
case $x_Q \gg x_\lambda$. The
double thick line stands for the system of source plus gluon cascade.}
\label{Figvariationhomogeneous}
\end{figure}
The first term is given by the extra particle amplitude squared times
the average of the operator $Q$ over the rest part of
the amplitude corresponding to the previous cut-off value $x_\lambda$
and recoil transverse momentum. It corresponds to the first diagram in
 Fig.\,\ref{Figvariationhomogeneous} and can be written symbolically
\beq
\label{D1symb}
\Delta_1 n_f(x_\lambda,q_\perp)\,=
 \,\langle \,p_B,x_\lambda, q_\perp-k_\perp,a,\sigma\,|
 a_{\lambda,\,c}\,Q
\,a^+_{\lambda,\,c}\,| \, p_B,x_\lambda,
 q^\prime_\perp-k_\perp,a^\prime,\sigma^\prime\,\rangle
\eq
or in explicit form
\begin{equation}
\label{D1}
\Delta_1 n_f(x_\lambda,q_\perp)\,=\,2N_c\,\frac{g^2}{(2\pi)^3}
\,\ln\frac{x_\lambda}{x_\lambda^\prime}\,
\int \frac{d^2k_\perp}{k_\perp^2} n_f(x_\lambda,q_\perp-k_\perp).
\end{equation}
Note that the previous equation is written in the case where the source
is in the adjoint representation. In the general case, $N_c$ should be
replaced by the Casimir of the corresponding representation.
 
It is important to note that
 in soft loops calculations gluons can be considered as
real, in accordance with the fact that there is almost no
difference between real and virtual massless soft particles.
Indeed, consider the loop where one single gluon is emitted and
reabsorbed, as illustrated in the Fig.\,\ref{Figvariationhomogeneous}.
The $\alpha$ integral which occur in the loops can be closed
around the pole of the emitted gluon propagator. For, using the same
technique which leads to the eikonal type expression (\ref{sv}), the
denominator of the integrand reads
$
(\alpha - i \delta)^2(\alpha \beta s - k_{\perp}^2 + i \delta)
$
for the first graph of Fig.\,\ref{Figvariationhomogeneous} or
$
(\alpha - i \delta)(\alpha \beta s - k_{\perp}^2 + i \delta)
$
for the second and third graphs. Both denominators leave
the singularities in $\alpha$ plane on the opposite side of the real axis.
The numerator of the gluon propagator in light-cone gauge reads
$$
d_{\mu\nu}(k)\,=\,-\varepsilon^\lambda_\mu(k)\, \varepsilon^\lambda_\nu(k)
\,-\,\frac{4k^2}{\beta^2s^2}\,p_{A\mu}\, p_{A\nu} \,.
$$
When performing the $\alpha$ integral, if among the two poles one choose
to close around the physical pole of the gluon,
the second term drops out since it cancels the $\alpha$ singularity in
the propagator. Thus, only the first term remains.
One then immediately gets
the soft universal vertex squared (\ref{universal}) for the first graph.

The second term in the variation arises due to virtual correction,
\begin{equation}
\label{D2}
\Delta_2 n_f(x_\lambda,q_\perp)\,=\,-2\ln\frac{x_\lambda}{x_\lambda^\prime}\,
\omega(q_\perp).
\end{equation}
It is illustrated by the second and third diagrams 
of Fig.\,\ref{Figvariationhomogeneous}. 

Differentiating with respect $x_\lambda$ we arrive at the evolution equation
$$
x_\lambda\frac{\partial}{\partial x_\lambda}n_f(x_\lambda,q_\perp)\,=\,
-2 \, N_c\frac{g^2}{(2\pi)^3}
\int \frac{d^2k_\perp}{(q-k)_\perp^2}\,n_f(x_\lambda,k_\perp)\,-\,
2\omega(q_\perp)\,n_f(x_\lambda,q_\perp)\,.
$$

In the case where $x_Q$ is smaller than the typical value of
$x_\lambda,$
there appears an additional term when the extra soft gluon operator is
contracted with the operator $Q,$ as shown by the additional fourth
diagram
in Fig.\,\ref{Figvariationinhomogeneous}.
\begin{figure}[htp]
\begin{picture}(20000,5000)
\global \Xone = 500                
\global \Yone = \Xone
\global \multiply \Yone by 2       
\global \Xtwo = 1300			 
\global \Xthree = \Xtwo
\negate \Xone
\global\advance \Xthree  by \Xone          
\negate \Xone
\global \Xfour = 300			   
\global \Xfive = 1100			   
\global \Ytwo = -250
\global \Ythree = -100		           
\global \Yfour = \Ythree
\global \multiply \Yfour by 2
\global \Xeight = 500		
\global \Yeight = 500
\THICKLINES
\drawline\fermion[\E\REG](\Xeight,\Yeight)[\Xtwo]
\THINLINES
\global\advance\fermionbacky by \Ythree
\drawloop\gluon[\N 5](\pbackx,\pbacky)
\put(\loopmidx,\fermionbacky){\circle{\Yone}}
\negate \Xone
\global\advance\loopmidx by \Xone			
\negate \fermionbackx
\global\advance\loopmidx by \fermionbackx
\Xseven = \loopmidx
\advance \Xseven by \Xtwo				
\negate \Xone
\THICKLINES
\drawline\fermion[\E\REG](\loopfrontx,\loopfronty)[\loopmidx]
\drawline\fermion[\W\REG](\gluonbackx,\gluonbacky)[\fermionlengthx]
\drawline\fermion[\E\REG](\gluonbackx,\gluonbacky)[\Xtwo]
\advance\fermionbacky by \Yfour
\drawline\fermion[\E\REG](\Xeight,\fermionbacky)[\Xseven]
\advance\fermionbackx by \Yone
\drawline\fermion[\E\REG](\fermionbackx,\fermionbacky)[\Xseven]
\THINLINES
\startphantom
\drawline\fermion[\E\REG](\pbackx,\pbacky)[\Xfour]
\stopphantom
\global \Ysix = \Yeight
\global \advance\Ysix \Ytwo
\put(\pbackx,100){$+$}
\startphantom
\drawline\fermion[\E\REG](\pbackx,\pbacky)[\Xfive]
\stopphantom
\THICKLINES
\drawline\fermion[\E\REG](\pbackx,\Yeight)[\Xtwo]
\Xseven = \fermionfrontx			
\THINLINES
\drawloop\gluon[\N 5](\pbackx,\pbacky)
\negate\loopfrontx
\global\advance\loopbackx by \loopfrontx
\THICKLINES
\drawline\fermion[\E\REG](\fermionbackx,\pbacky)[\loopbackx]
\drawline\fermion[\E\REG](\fermionbackx,\fermionbacky)[\Xthree]
\negate \fermionbackx
\advance \Xseven by \fermionbackx
\negate \fermionbackx
\negate \Xseven			
\global\advance\fermionbackx by \Xone
\global\advance\fermionbacky by \Ythree
\THINLINES
\put(\fermionbackx,\fermionbacky){\circle{\Yone}}
\THICKLINES
\global\advance\fermionbackx by \Xone
\negate\Ythree
\advance\fermionbacky by \Ythree
\negate\Ythree
\drawline\fermion[\E\REG](\fermionbackx,\fermionbacky)[\Xthree]
\Yseven = \fermionbackx
\advance \fermionbacky by \Yfour
\drawline\fermion[\W\REG](\fermionbackx,\fermionbacky)[\Xthree]
\negate \Yone
\advance \fermionbackx by \Yone
\negate \Yone
\drawline\fermion[\W\REG](\fermionbackx,\fermionbacky)[\Xseven]
\THINLINES
\pbackx = \Yseven
\startphantom
\drawline\fermion[\E\REG](\pbackx,\pbacky)[\Xfour]
\stopphantom
\global \Ysix = \Yeight
\global \advance\Ysix \Ytwo					

\put(\pbackx,100){$+$}
\startphantom
\drawline\fermion[\E\REG](\pbackx,\pbacky)[\Xfive]
\stopphantom
\THICKLINES
\drawline\fermion[\E\REG](\pbackx,\Yeight)[\Xthree]
\Yseven = \fermionfrontx			
\global\advance\fermionbackx by \Xone
\global\advance\fermionbacky by \Ythree
\THINLINES
\put(\fermionbackx,\fermionbacky){\circle{\Yone}}
\THICKLINES
\global\advance\fermionbackx by \Xone
\Xseven = \fermionbackx					
\negate\Ythree
\advance\fermionbacky by \Ythree
\negate\Ythree
\drawline\fermion[\E\REG](\fermionbackx,\fermionbacky)[\Xthree]
\THINLINES
\drawloop\gluon[\N 5](\pbackx,\pbacky)
\negate\loopfrontx
\global\advance\loopbackx by \loopfrontx
\THICKLINES
\drawline\fermion[\E\REG](\fermionbackx,\fermionbacky)[\loopbackx]
\drawline\fermion[\E\REG](\fermionbackx,\fermionbacky)[\Xthree]
\negate \fermionbackx
\advance \Xseven by \fermionbackx
\negate \Xseven
\advance \fermionbacky by \Yfour
\drawline\fermion[\E\REG](\Yseven,\fermionbacky)[\Xthree]
\advance \fermionbackx by \Yone
\drawline\fermion[\E\REG](\fermionbackx,\fermionbacky)[\Xseven]
\THINLINES
\startphantom
\drawline\fermion[\E\REG](\pbackx,\pbacky)[\Xfour]
\stopphantom
\global \Ysix = \Yeight
\global \advance\Ysix \Ytwo					
\put(\pbackx,100){$+$}
\startphantom
\drawline\fermion[\E\REG](\pbackx,\pbacky)[\Xfive]
\stopphantom
\THICKLINES
\drawline\fermion[\E\REG](\pbackx,\Yeight)[\Xtwo]
\Yseven = \fermionfrontx					
\THINLINES
\drawloop\gluon[\N 3](\pbackx,\pbacky)
\global\advance \gluonbackx by \Xone
\put(\gluonbackx,\gluonbacky){\circle{\Yone}}
\global\advance \gluonbackx by \Xone
\drawloop\gluon[\E 3](\gluonbackx,\gluonbacky)
\global\negate\fermionbackx
\global\advance\gluonbackx by \fermionbackx
\global\negate\fermionbackx
\THICKLINES
\drawline\fermion[\E\REG](\fermionbackx,\fermionbacky)[\gluonbackx]
\drawline\fermion[\E\REG](\fermionbackx,\fermionbacky)[\Xtwo]
\negate \fermionbackx
\advance \Yseven by \fermionbackx
\negate \Yseven
\negate \fermionbackx
\advance \fermionbacky by \Yfour
\drawline\fermion[\W\REG](\fermionbackx,\fermionbacky)[\Yseven]
\THINLINES

\end{picture}
\vskip 0.5cm
\caption{Variation of the density with respect to the cut-off, in the
case $x_Q \ll x_\lambda.$}
\label{Figvariationinhomogeneous}
\end{figure}
It gives the following contribution to the variation of $n_f:$
\beq
\label{defDelta3}
\Delta_3n_f(x_\lambda,q_\perp) = 2 \, N_c \,  \frac{g^2}{(2 \, \pi)^3}
 f(x_\lambda,q_\perp)
 \ln \frac{x_\lambda}{x'_\lambda}\,.
\eq
The full inhomogeneous equation thus reads
\bea
\label{inhomo}
&&x_\lambda\frac{\partial}{\partial x_\lambda}n_f(x_\lambda,q_\perp)\,=\,2 \, N_c \,  \frac{g^2}{(2 \, \pi)^3}
 \int \frac{d^2 l_\perp}{l^2_\perp} \, 
 \ln \frac{x_\lambda}{x'_\lambda} \nonumber \\
&&
-2 \, N_c\frac{g^2}{(2\pi)^3}
\int \frac{d^2k_\perp}{(q-k)_\perp^2}\,n_f(x_\lambda,k_\perp)\,-\,
2\omega(q_\perp)\,n_f(x_\lambda,q_\perp)\,.
\eea
If one is interested in the density number of gluon of momentum $r_\perp,$ the weight
function $f$ in Eq.(\ref{defQ}) should be chosen as
\beq
\label{defnumber}
f(x,l_\perp)=\frac{1}{2}\, \delta^2(l_\perp-r_\perp)\,,
\eq
the $1/2$ factor being related to the average over gluon transverse polarization.

In that case, Eq.(\ref{inhomo}) turns into the BFKL equation (for $t=0$)
\bea
\label{BFKL}
&&x_\lambda\frac{\partial}{\partial x_\lambda}G(\cut,\, r_\perp,\,q_\perp)\,=- \, N_c \,  \frac{g^2}{(2 \, \pi)^3}
 \frac{\delta^2(q_\perp-r_\perp)}{r^2_\perp} \nonumber \\
&&
-2 \, N_c\frac{g^2}{(2\pi)^3}
\int \frac{d^2k_\perp}{(q-k)_\perp^2}\,G(x_\lambda,\,
 r_\perp,\, k_\perp)\,-\,
2\omega(q_\perp)\,G(x_\lambda,\, r_\perp,\, q_\perp)\,.
\eea
The inhomogeneous term, which is of lowest order, can be
interpreted as the initial gluon contribution in the absence of cascade.
Since we consider here gluons emitted from a source, 
it 
contains a factor $N_c \, g^2/(2 \pi)^3,$ multipliying $1/r_\perp^2$ which
 is reminiscent of the two $t-$channel gluon propagators (one being compensated
 by polarization contribution after angular averaging of $k_\perp$).

The equation Eq.(\ref{BFKL}) looks like a Bethe-Salpeter equation
 with a kernel $K$ acting
from below, which is illustrated in Fig.\,\ref{FigBFKL}.
\begin{figure}
\begin{center}               
\begin{picture}(24000,12000)   
\global\Xone=2980 
\global\Yone=100 
\global\Xtwo=\Xone
\global\multroothalf\Xtwo
\drawline\fermion[\E\REG](0,0)[\Xtwo]
\global\advance\pfrontx by -4000
\put(\pfrontx,-300){$x_\lambda\frac{\partial}{\partial x_\lambda}$}
\drawline\gluon[\N\FLIPPED](\pbackx,\pbacky)[3]
\global\Xthree=\gluonbackx
\global\Ythree=\gluonbacky
\drawline\fermion[\E\REG](\fermionbackx,\fermionbacky)[\Xtwo]
\drawline\gluon[\N\REG](\pbackx,\pbacky)[3]
\drawline\fermion[\E\REG](\fermionbackx,\fermionbacky)[\Xtwo]
\startphantom
\drawline\fermion[\NW\REG](\gluonbackx,\gluonbacky)[\Xone]
\stopphantom
\drawline\gluon[\N\FLIPPED](\pbackx,\pbacky)[2]
\startphantom
\drawline\fermion[\NE\REG](\Xthree,\Ythree)[\Xone]
\stopphantom
\global\Xthree=\pmidx
\global\Ythree=\pmidy
\drawline\gluon[\N\REG](\pbackx,\pbacky)[2]
\global\Ytwo=\Xone
\global\multiply\Ytwo by 115
\global\divide\Ytwo by 109
\THICKLINES
\put(\Xthree,\Ythree){\circle{3001}}   
\THINLINES
\global\advance\Xthree by -300
\global\advance\Ythree by -300
\put(\Xthree,\Ythree){$G$}
\global\multiply\Xtwo by 3
\global\negate\Yone
\drawline\fermion[\E\REG](0,\Yone)[\Xtwo]
\global\divide\Xtwo by 3
\startphantom
\drawline\fermion[\E\REG](\pbackx,\pbacky)[\Xtwo]
\stopphantom
\global\negate\Yone
\global\advance\pfronty by \Yone
\startphantom
\drawline\fermion[\E\REG](\pfrontx,\pfronty)[5000]
\stopphantom
\put(8450,-300){$=$}
\drawline\fermion[\E\REG](\pbackx,\pbacky)[\Xtwo]
\global\Xthree=\pfrontx
\global\Ythree=\pfronty
\global\negate\Yone
\global\advance\Ythree by \Yone
\drawline\gluon[\N\FLIPPED](\pbackx,\pbacky)[8]
\drawline\fermion[\E\REG](\fermionbackx,\fermionbacky)[\Xtwo]
\drawline\gluon[\N\REG](\pbackx,\pbacky)[8]
\drawline\fermion[\E\REG](\fermionbackx,\fermionbacky)[\Xtwo]
\global\multiply\Xtwo by 3
\drawline\fermion[\E\REG](\Xthree,\Ythree)[\Xtwo]
\global\divide\Xtwo by 3
\startphantom
\drawline\fermion[\E\REG](\pbackx,\pbacky)[\Xtwo]
\stopphantom
\global\negate\Yone
\global\advance\pfronty by \Yone
\startphantom
\drawline\fermion[\E\REG](\pfrontx,\pfronty)[5000]
\stopphantom
\put(19800,-300){$+$}
\drawline\fermion[\E\REG](\pbackx,\pbacky)[\Xtwo]
\drawline\gluon[\N\FLIPPED](\pbackx,\pbacky)[2]
\drawline\fermion[\E\REG](\fermionbackx,\fermionbacky)[\Xtwo]
\drawline\gluon[\N\REG](\pbackx,\pbacky)[2]
\drawline\fermion[\E\REG](\fermionbackx,\fermionbacky)[\Xtwo]
\global\multiply\Xtwo by 3
\global\negate\Yone
\global\advance\fermionbacky by \Yone
\drawline\fermion[\W\REG](\fermionbackx,\fermionbacky)[\Xtwo]
\global\negate\Yone
\global\divide\Xtwo by 3
\THICKLINES
\drawline\fermion[\W\REG](\gluonbackx,\gluonbacky)[\Xtwo]
\THINLINES
\global\divide\Xtwo by 2
\THICKLINES
\drawline\fermion[\N\REG](\pbackx,\pbacky)[2000]
\global\multiply\Xtwo by 2
\THINLINES
\drawline\gluon[\N\FLIPPED](\pbackx,\pbacky)[2]
\global\Xthree=\gluonfrontx
\global\Ythree=\gluonfronty
\startphantom
\drawline\fermion[\NE\REG](\gluonbackx,\gluonbacky)[\Xone]
\stopphantom
\drawline\gluon[\N\REG](\pbackx,\pbacky)[2]
\THICKLINES
\drawline\fermion[\E\REG](\Xthree,\Ythree)[\Xtwo]
\THINLINES
\drawline\gluon[\N\REG](\pbackx,\pbacky)[2]
\global\Xthree=\gluonfrontx
\global\Ythree=\gluonfronty
\startphantom
\drawline\fermion[\NW\REG](\gluonbackx,\gluonbacky)[\Xone]
\stopphantom
\THICKLINES
\put(\pmidx,\pmidy){\circle{3001}}     
\THINLINES
\global\advance\pmidx by -300
\global\advance\pmidy by -300
\put(\pmidx,\pmidy){$G$}
\drawline\gluon[\N\FLIPPED](\pbackx,\pbacky)[2]
\global\divide\Xtwo by 2
\THICKLINES
\drawline\fermion[\S\REG](\Xthree,\Ythree)[2000]
\global\multiply\Xtwo by 2
\put(25400,3000){$K$}

\end{picture}
\end{center}
\caption{BFKL equation for the gluon density in a cascade. The double
line stands for the source.}
\label{FigBFKL}
\end{figure}

Actually the density $G$ is only a function of $p_\perp - r_\perp$ due
to translational invariance. It means that Eq.(\ref{BFKL}) could 
equally be written as a Bethe-Salpeter equation with a kernel acting from
above, by just an exchange between of variables $r_\perp$ and $p_\perp.$

A natural physical value for the IR cut-off $\cut$ is $\cut \sim
\mu^2/s$ where $\mu^2$ is some typical scale for tranverse momentum.
This follows from the Regge kinematics (\ref{kin}).

Using Eqs.(\ref{defQ}) and (\ref{defnumber}), one can 
finally obtain the unintegrated parton density from $G$
as
\bea
\label{unintegrated}
&& g(\cut, r_\perp, \, q_\perp) \, =\, \frac{1}{2 \cut} \langle \,p_B,\, \cut,
q_\perp,a,\sigma\,|\, \frac{1}{2} \sum_{\lambda,c} a^+_{\sigma,c}\,
a_{\sigma,c}
|\,p_B,\, \cut, q_\perp^\prime,a^\prime,\sigma^\prime\,\rangle\,
\nonumber \\
&& =\, -\frac{\partial}{\partial \cut} G(\cut,\, r_\perp, \, q_\perp)\,.
\eea
In this formula, $r_\perp$ is the tranverse momentum at which parton
density is measured while $q_\perp$ is the total tranverse momentum of
the cascade plus source.

\section{Computation of $\omega$}
\label{computationomega}

\subsection{Evaluation of $\omega$ based on the gluon cascade universality}
\label{universality}

The expression (\ref{D1}) provides a simple way to find the function
$\omega(q_\perp)$ based on universality. 
From the point of view of emitted gluon, the whole system corresponding to the previous
cut-off value plays the role of the source, and the $z$-factor is prescribed to this whole system.
On the other hand, this whole system looks like a gluon (in the case where bare source is 
gluon like). Thus, the $z$-factor for the whole system and for a single
gluon should  have the same functional form.
What we need
is the virtual correction for the whole system, but we shall calculate
instead
the
$z$-factor of the dressed emitted gluon. The inclusion of $z$-factor 
corresponds
to
the replacement $a_{\lambda,\sigma} \to  z^{1/2} \, a_{R
\,\lambda,\sigma}$
in the expression (\ref{D1symb}).
With $z$-factor included the formula
(\ref{D1}) reads
\begin{eqnarray}
\label{Dpi}
\Delta_1 n_f(x_\lambda,q_\perp)\,&=&\,2N_c\,\frac{g^2}{(2\pi)^3}
\ln\frac{x_\lambda}{x_\lambda-\delta x_\lambda} \\
&\times&\,\int \frac{d^2p_\perp}{p_\perp^2}
z_{[x_\lambda-\delta x_\lambda,x_\lambda]}(p_\perp)\,
n_f(x_\lambda,q_\perp-p_\perp). \nonumber
\end{eqnarray}
On the other hand there is a dual description of the same dressed gluon as a
composite cascade state spread over the interval
$[x_\lambda-\delta x_\lambda,x_\lambda]$ and carrying transverse momentum
$p_\perp$. The two particles component of this state is given by the amplitude
to emit two soft gluons off the source (which is the whole system corresponding to the previous
cut-off value),
$$
|\,p_B,\,q_\perp,\,x_\lambda-\delta x_\lambda,\,a\,\sigma\,\rangle_2\,=\,
igf_{ac_1d}\,igf_{dc_2b}\!\!\!\!
\int\limits_{x_\lambda-\delta x_\lambda}^{x_\lambda}\!\!
\frac{dx}{2x}\!\! \int\limits_{x_\lambda-\delta x_\lambda}^{x}\!\!
\frac{d\beta}{2\beta}\,
\int\frac{d^2k_\perp}{(2\pi)^{3/2}}\,
2\frac{k_\perp\cdot\varepsilon_{\sigma_2}}{k_\perp^2}
$$
$$
\times\frac 1{(2\pi)^{3/2}}\,
2\frac{(p-k)_\perp\cdot\varepsilon_{\sigma_1}}{(p-k)_\perp^2}\,
a_{\sigma_2c_2}^+(\beta,k_\perp)\,a_{\sigma_1c_1}^+(x,p_\perp-k_\perp)
|\,p_B,\,q_\perp-p_\perp,\,x_\lambda,\,b,\, \sigma \,\rangle,
$$
the gluons being ordered with respect to their longitudinal momenta,
$$
x_\lambda-\delta x_\lambda\,<\,\beta\,<\, x\, <\,x_\lambda,
$$
which is reflected in the integration limits.

Averaging the operator with this amplitude we arrive at the following
expression for the gluon density variation
$$
\Delta_1 n_f(x_\lambda,q_\perp)\,=\,2N_c\,\frac{g^2}{(2\pi)^3}\,
\frac{1}{2} \ln^2\frac{x_\lambda}{x_\lambda-\delta x_\lambda}\,
\int \frac{d^2p_\perp}{p_\perp^2}\,n_f(x_\lambda,q_\perp-p_\perp)
$$
$$
\times\,2N_c\,\frac{g^2}{(2\pi)^3}\,\int d^2k_\perp\,
\frac{p_\perp^2}{k_\perp^2(p-k)_\perp^2}.
$$
Note that the double integration in $x$ and $\beta$ results into $1/2
\, \ln^2 \frac{x_\lambda}{x_\lambda-\delta x_\lambda}.$ 
This factor $1/2$ reflects the Bose symmetry of the
two gluons system.
Recalling Eq.(\ref{zomega}) and Eq.(\ref{Dpi}) we get
\begin{equation}
\label{omega_2}
2\,\omega(p_\perp)\,=\,N_c\,\frac{g^2}{(2\pi)^3}\,\int d^2k_\perp\,
\frac{p_\perp^2}{k_\perp^2(p-k)_\perp^2}.
\end{equation}

\subsection{Direct calculation}
\label{direct}

The formula (\ref{omega_2}) can be compared with a direct calculation
of the polarization operator. The exact one loop result in the dimensional
regularization ($D=2+2\epsilon$) and axial gauge is
\begin{equation}
\label{oneloop}
\pi(p_\perp)\,=\,-2N_c\,\frac{g^2}{8\pi^2}\,\Gamma\biggl(1-\frac D2\biggr)\,
\int_0^1 d\beta\,\bigl[\beta (1-\beta)\,p_\perp^2/4\pi\bigr]^{\frac D2-1}
(1-\beta) \,K(\beta),
\end{equation}
where
$$
K(\beta)\,=\,\frac{\beta}{1-\beta}\,+\,\frac{1-\beta}{\beta}\,+\,
\beta\,(1-\beta)
$$
is the DGLAP kernel. Gluon self-energy is known to coincide in the axial gauge
with a Sudakov form factor whose double logarithmic behavior
originates from the product of transverse and longitudinal
divergencies. The latter one comes about when one gluon momentum in the loop
becomes soft, $\beta \to 0$. To separate the longitudinal logarithms
needed in order to find the function $\omega(p_\perp)$ we keep only
the singular piece in $K(\beta)$ and cut the integral in (\ref{oneloop})
at $\beta=x_\lambda$, using dimensional regularization only for
tranverse divergency, which turns into
\begin{equation}
\label{omega_n}
\pi_1(x_\lambda,p_\perp^2)\,=\,-2N_c\,\frac{g^2}{8\pi^2}\,
\Gamma\biggl(1-\frac D2\biggr)\,
\bigl[p_\perp^2/4\pi\bigr]^{\frac D2-1}\,\ln\frac 1{x_\lambda}\,=\,
2\,\omega(p_\perp)\,\ln\frac 1{x_\lambda}.
\end{equation}
This results into
$$
2\,\omega(p_\perp)\,=\,2\,N_c\,\frac{g^2}{8\pi^2}\,\biggl[\frac 1\epsilon\,+\,
\ln \frac{p_\perp^2}{4\pi}\,+\,\gamma_E\biggr]\,+\,O(\epsilon),
$$
while dimensionally regularized integral (\ref{omega_2}) gives
\begin{eqnarray}
2\omega(q_\perp)\,&=&\,
N_c\,\frac{g^2}{8\pi^2}\,\bigl[q_\perp^2/4\pi\bigr]^{\frac D2-1}
\Gamma\biggl(2-\frac D2\biggr)\,\frac{\Gamma^2(\frac D2-1)}{\Gamma(D-2)}
\nonumber \\
&=&\,2N_c\,\frac{g^2}{8\pi^2}\,\biggl[\frac 1\epsilon\,+\,
\ln \frac{q_\perp^2}{4\pi}\,+\,\gamma_E\biggr]\,+\,O(\epsilon)\,. \nonumber
\end{eqnarray}
Thus, although functions $\omega(p_\perp)$ defined by eq.(\ref{omega_2})
and eq.(\ref{omega_n}) look different their nonvanishing parts are equal.

Further, if we consider the renormalization of a heavy color source of mass
$M$ due to
classical current emission, the divergent part of the source
self energy amounts to the same expression
\beq
\label{piM}
\pi_M(x_\lambda,p_\perp^2)\,=\,-2N_c\,\frac{g^2}{8\pi^2}\,
\Gamma\biggl(1-\frac n2\biggr)\,
\left[\frac{M^2-p^2}{4\pi}\right]^{\frac n2-1}\,\ln\frac 1{x_\lambda},
\eq
with source mass $M$ and virtuality $M^2-p^2\simeq p_\perp^2$ regardless
of an elementary or composite nature of the source. 

\section{QED cascade}
\label{QEDcascade}

The above treatment
can be applied to electrodynamics.
Generally speaking, the wave function of the source, which could be
any charged object, for instance an electron, is made of the bare source
itself surrounded by a cloud of soft photons.
The operator $Q$ can probe either the source or photon component of 
the wave function. If we introduce a cut-off for the longitudinal
momentum of the soft photons, we can study the effect of changing this
cut-off as we did for the QCD case. We get the same three type of
contributions as illustrated in Fig.\,\ref{Figvariationinhomogeneous},
where the wavy lines now represent photons.

 The $z$-factor is related in
this case to the renormalization of the source, and can be obtained 
from Eq.(\ref{piM}).  One needs only to replace
$g$ by the electric charge $Z \, e$ and put $N_c=1$. The real
contribution
is only due to the emission off the source line, since photon has no
charge.
In the corresponding Wilson line (\ref{wilson}), the ordering of the
fields is not important since they commute at equal light cone time.
It is related to the fact that in QED, the soft cascade structure
does not assume any additional 
$\beta$ ordering (\ref{betaordering}).

The equation 
satisfied by the corresponding $n_f$ is the same as Eq.(\ref{inhomo}),
after performing the replacement $g^2 N_c \to Z^2 \, e^2 \,.$
In the case of photon density, one gets
\bea
\label{BFKLQED}
&&x_\lambda\frac{\partial}{\partial
 x_\lambda}G_\gamma(\cut,\, r_\perp,\,q_\perp)\,=-  \,  \frac{Z^2 \, e^2}{(2 \, \pi)^3}
 \frac{\delta^2(q_\perp-r_\perp)}{r^2_\perp} \nonumber \\
&&
-2 \frac{Z^2 \,e^2}{(2\pi)^3}
\int \frac{d^2k_\perp}{(q-k)_\perp^2}\,G_\gamma(\cut,\,r_\perp,\,k_\perp)\,-\,
2\omega(q_\perp)\,G_\gamma(\cut,\,r_\perp,\,q_\perp)\,.
\eea
This equation has just the same form than the BFKL equation (\ref{BFKL})
 for QCD.
Let us stress once more that contrarily to the QCD case,
 $\omega(q_\perp)$ corresponds to the
 reggeization
of the source and not of the photon, which is of different nature.
In QCD, in the case where the source is gluon like, $\omega(q_\perp)$
can be associated either to the source or to the cascade.
We have used this property to get the expression for  $\omega(q_\perp).$
From the point of view of source dressing, it is obtained through
the one loop gluon self energy (see Eq.(\ref{omega_n})). Equivalently,
it is obtained through
the normalisation of the wave function (see Eq.(\ref{omega_2})).

\section{Cascade wave function}
\label{cascadewave}

Let us now turn back to the amplitude (\ref{prodPhi}). Despite the
symbolic character of the infinite ordered product it can be presented
in a closed form. Consider to this end the amplitude to emit $n$ quasi-real
gluons with momenta $x_n,q_{\perp_1},\ldots,x_1,q_{\perp_1}$,
$x_1>x_2>\ldots>x_n$. It is given by the matrix element
\bea
\label{matrix}
&&
\Gamma_{\lambda_n,\,c_n,\,\ldots,\,\lambda_1,\,c_1;\,\sigma',\,\sigma;\,a',\,a
}
(x_n,q_{\perp n},\ldots,x_1,q_{\perp 1};\,q_\perp',\, q_\perp)\\
&& =
\langle\,0\,|\,b_{\sigma',a^\prime}(p_B,q_\perp^\prime)\,
a_{\lambda_n,c_n}(x_n,q_{\perp_n})\,\cdots\,
a_{\lambda_1,c_1}(x_1,q_{\perp_1})\,\Phi^+(x_\lambda)\,
b_{\sigma,a}^+(p_B,q_\perp)\,|\,0\,\rangle \,. 
\nonumber
\eea
Expanding the brackets, the product in the $\Phi^+(x_\lambda)$ operator
can be reorganized as
$$
\Phi^+(x_\lambda)\,=\,\prod\limits_{\cut}^{x_n}
\left[1 - \omega(\hat P)\,\frac{\step}{x}\right]\,
\int_{x_n-\step}^{x_n}\,d\beta_n\,\int d^2k_{\perp_n}\,
V^+(\beta_n,\,k_{\perp_n})\,
$$
$$
\times\,\prod\limits_{x_{n-1}}^{x_n}
\left[1 - \omega(\hat P)\,\frac{\step}{x}\right]\,\cdots\,
\prod\limits_{x_1}^{x_2}
\left[1 - \omega(\hat P)\,\frac{\step}{x}\right]\,
$$
$$
\times\,
\int_{x_1-\step}^{x_1}\,d\beta_1\,\int d^2 k_{\perp_1}\,
V^+(\beta_1,\,k_{\perp_1})\,
\prod\limits_{x_1}^{1}
\left[1 - \omega(\hat P)\,\frac{\step}{x}\right] + \cdots \,.
$$
The terms which are not explicitly written gives zero contribution to
the matrix element (\ref{matrix}) when performing the contractions 
with $a$ operators.
Using the fact that
$$
\prod\limits_{x_2}^{x_1}
\left[1 - \omega(\hat P)\,\frac{\step}{x}\right]\,=\,
\exp\left\{-\int_{x_2}^{x_1}\frac{d\beta}{\beta}\omega(\hat P)\right\}
$$
and that the operator $\hat P$ results into the total
momentum of the state occuring to the right, we get for the amplitude
$$
\Gamma_{\lambda_n,\,c_n,\,\ldots,\,\lambda_1,\,c_1;\,\sigma',\,\sigma;\,a',\,a
}
(x_1,q_{\perp 1},\ldots,x_n,q_{\perp n};\,q_\perp',\, q_\perp)
$$
$$
=\, \left(\frac{\cut}{x_n}\right)^
{\omega(q_{\perp_n}+\cdots+q_{\perp_1}+q_\perp)}\Gamma^{\lambda_n}_{c_n}(q_{\perp 1})
\left(\frac{x_n}{x_{n-1}}\right)^
{\omega(q_{\perp_{n-1}}+\cdots+q_{\perp_1}+q_\perp)}\Gamma^{\lambda_{n-1}}_{c_{n-1}}(q_{\perp_{n-1}})
\cdots\,
$$
$$
\times\,\Gamma^{\lambda_1}_{c_1}(q_{\perp_1})
\left(\frac{x_1}{1}\right)^{\omega(q_\perp)}
\langle\,0\,| \, b_{\sigma',a^\prime}(p_B,q_\perp^\prime)\, 
b_{\sigma,a}^+(p_B,q_\perp)\,|\,0\,\rangle 
.
$$
This formula, in the LLA approximation, provides all information on the
cascade wave function. It resembles multi-reggeon formula but it
includes
only soft vertex. 

\section{Conclusion}

This paper is based
on the ordering of longitudinal variables in soft cascades,
which means energy ordering of emitted particles.
It differs from
collinear approximation which implies angular 
ordering \cite{Dokshitzer:1991wu}. 
In this sense, this treatment describes the dynamics 
responsible for BFKL evolution, while the collinear one
is related to the conventional partonic DGLAP picture.
We have shown explicitly that the infrared evolution equation for the parton
density in the soft cascade reproduces forward BFKL equation. 

 In both BFKL and DGLAP equations the evolution is written with respect to a
cut-off. DGLAP  involves a transverse cut-off $Q^2$ due to collinear
singularities, while in our case the cut-off is longitudinal, due to soft
singularities.

In DGLAP case the unintegrated parton distributions, which depend both
on the longitudinal and tranverse momenta, are related to the derivative
of the structure functions with respect to the
transverse cut-off.
Similarly we have to consider the derivative  with respect to
$x_\lambda$ of the function
$G(x_\lambda,r_\perp,p_\perp),$ which satisfies BFKL equation,
in order to get the unintegrated distribution.

Our approach is based on a kind of duality in the description
of a gluon at high energy.
From the viewpoint of the soft vertex
the gluon state can be treated as a single particle with given momentum and
color. On the other hand the gluon has internal structure, a
cloud of many soft gluons surrounding it. It looks like a composite
object with a nontrivial wavefunction. Hence a rather simple interpretation
of reggeization as a manifestation of soft cascades structure appears in this
picture. The trajectory $\omega(q_\perp)$ is nothing more than the
gluon $Z$ factor in the axial gauge. Both in QCD and QED, this function 
is associated with the charged source self energy.
In DGLAP case the divergencies  occurring at $x \to 1$ are regulated
by the $1/(x-1)_+$ prescription, corresponding to UV $Z$-factors
due to UV renormalization of parton wave functions.
In BFKL case the IR divergencies for $k_\perp \to 0$ disappear because
of virtual corrections which can be treated as IR renormalization of
the gluon wave function. The equation (\ref{BFKL}) can be wieved as describing  $2$
reggeized gluons, or, equivalently, $2$ gluon soft cascades.

The extension of this approach for non forward case and for Generalized
Leading Log Approximation (\cite{bBJKP,jBJKP,kpBJKP}) will be presented
in a forthcoming paper.
 It would be also interesting to study more complicated
structures like for example the triple Pomeron vertex.

\section*{Acknowledgements}

We thank G. Korchemsky, L. Szymanowski and O.V. Teryaev
for comments, and A.S is grateful to M. Ryskin
for many useful discussions.

We thank
J. Bartels  and the II. Instit\"ut f\"ur Theoretische Physik at DESY for support at the
beginning of this work. A.S acknowledge for support from the IPN, LPTMS
and LPT (Orsay).
S.W thanks the Alexander von Humboldt Foundation for support when this
work was at an early stage.
A.S thanks LPNHE for hospitality, he is grateful for a LPTHE-Steklov Institute
Visiting Fellowship agreement, for a CNRS Associated Research position
and for a NATO grant.

\end{document}